\documentclass[10pt,prb,aps,showpacs,twocolumn]{revtex4}
\usepackage[caption=false]{subfig}

\usepackage{amssymb}
\usepackage{amsmath}
\usepackage{commath}
\usepackage{graphicx,bm}
\usepackage{verbatim}
\usepackage[usenames]{xcolor}
\usepackage{pdfpages}
\usepackage{hyperref}
\usepackage[normalem]{ulem}

\def\be#1\ee{\begin{equation}#1\end{equation}}
\def\ba#1\ea{\begin{align}#1\end{align}}

\newcommand{\eqqref}[1]{Eq.\,\eqref{#1}}	
\newcommand{\figgref}[1]{Fig.\,\ref{#1}}

\newcommand{\PRB}[3]{Phys.\,Rev.\,B {\bf #1}, #2 (#3).}
\newcommand{\PRD}[3]{Phys.\,Rev.\,D {\bf #1}, #2 (#3).}
\newcommand{\PRL}[3]{Phys.\,Rev.\,Lett. {\bf #1}, #2 (#3).}

\newcommand{\bk}	{{\bf k}}

\newcommand{\bq}	{{\bf q}}

\newcommand{\bt}    {{\bf t}}

\newcommand{\eps}	{\epsilon}

\newcommand{\s}	    {\sigma}
\newcommand{\w}	    {\omega}

\newcommand{\NFD}[1]			   {N_{\text{F}}(#1)}

\newcommand{\mus}{\mu_{\s}}

\newcommand{\limhomg}{\lim_{\substack{q\rightarrow0 \\ \w\rightarrow0}}}
\newcommand{\tpara}{t_{\parallel}}

\begin{document}
\title{Anisotropic chiral magnetic effect from tilted Weyl cones}
\author{E.C.I. van der Wurff}
\email{e.c.i.vanderwurff@uu.nl}
\author{H.T.C. Stoof}
\affiliation{Institute for Theoretical Physics and Center for Extreme Matter and Emergent Phenomena, Utrecht University,  Princetonplein 5, 3584 CC Utrecht, The Netherlands}
\date{\today}
\begin{abstract}
We determine the antisymmetric current-current response for a pair of (type-I) tilted Weyl cones with opposite chirality. We find that the dynamical chiral magnetic effect depends on the magnitude of the tilt and on the angle between the tilting direction and the wave vector of the magnetic field. Additionally, the chiral magnetic effect is shown to be closely related to the presence of an intrinsic anomalous Hall effect with a current perpendicular to the tilting direction and the electric field. We investigate the nonanalytic long-wavelength limit of the corresponding transport coefficients.
\end{abstract}

\maketitle
\vspace{-.2cm}
\textit{Introduction.---} In classical electrodynamics, magnetic fields always induce currents that are perpendicular to the magnetic field direction due to the Lorentz force. However, in quantum electrodynamics, a current can also be generated in the same direction as the magnetic field. This was first realized for massless fermions in particle physics \cite{Ninomiya,Kharzeev}. It is a consequence of the fact that quantum mechanically a magnetic field quenches the kinetic energy perpendicular to its direction and also spin polarizes the lowest Landau level. As a result massless fermions only obtain a drift velocity along the magnetic field with an opposite sign for opposite chiralities. Inducing an imbalance between the two chiral species then gives a net current along the magnetic field direction known now as the chiral magnetic effect (CME).

Massless chiral fermions also occur as low-energy quasiparticles in the recently discovered Weyl (semi)metals \cite{Hasan1,Soljacic,Hasan2,Ding1,Ding2}. These quasiparticles do not move at the speed of light, as in elementary-particle physics, but rather at the Fermi velocity. Additionally, the effective Weyl cones with different chirality are in a real material always connected by the full bandstructure and hence electrons can be transported from one cone to another by applying both an electric and a magnetic field \cite{Burkov1}. In particle physics the same phenomenon occurs due to the breaking of chiral symmetry by quantum corrections. This breaking of chiral symmetry due to the renormalization of ultraviolet divergencies is called a chiral anomaly and causes the difference between the numbers of particles with positive and negative chirality to be no longer conserved \cite{Adler,Jackiw,Bardarson1}.

The main difference with particle physics is that Lorentz invariance is not enforced in a condensed-matter material. This gives, besides a velocity that is smaller than the speed of light, also the possibility that Weyl nodes are separated in energy-momentum space. Splitting them in the momentum direction gives rise to a topological anomalous Hall effect \cite{Burkov1}, whereas splitting them in the energy direction is exactly the situation of most interest for the CME \cite{Ninomiya,Burkov2}. Indirect measurements of the chiral magnetic effect have recently been made by the observation of a negative magnetoresistance \cite{Spivak,Valla,Hasan0,Bardarson2}. Another interesting possibility is tilting the Weyl cones, meaning that the slope of the dispersion relation is not the same in opposite directions \cite{Goerbig, Bergholtz}. Materials that exhibit such tilted Weyl cones are of type I if the tilt is relatively small and of type II if the cones are overtilted such that the electron and hole dispersions intersect the energy plane of the Weyl node itself \cite{Bernevig,Tchoumakov1,Zhang}. Moreover, the tilt is affected and can even be generated by disorder and interaction effects \cite{Bergholtz2,Stoof,Fritz1}. It is thus of considerable interest to investigate what such a tilt does to the chiral magnetic conductivity of a Weyl (semi)metal.

\begin{figure}[t!]
\includegraphics[trim={0cm 0cm 0cm 0cm},clip,scale=1.48]{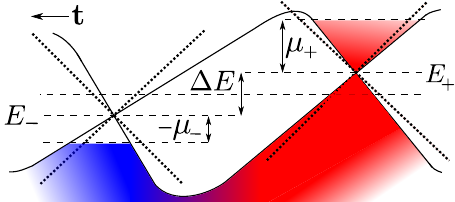} 
\vspace{-.3cm}
\caption{Illustration of a band structure with two imbalanced Weyl cones at chiral chemical potentials $\mu_{\pm}$ and node energies $E_{\pm}$, both tilted in the same direction $\bt$.} \label{fig:FrontPage}
\vspace{-.3cm}
\end{figure}
The chiral magnetic conductivity is in principle a function of the wavenumber and frequency of the applied magnetic field \cite{Explain0}. When calculating the long-wavelength limit, the order of limits is crucial and we need to distinguish the case in which the Weyl nodes are located at the same energy and the case in which they are not \cite{Burkov3,ChangYang,Moore,Pesin1}. Only when the chiral imbalance of the two Weyl nodes is exactly opposite to their energy separation, is there a vanishing current in the static limit \cite{Burkov2,Franz,Nazarov}.

\begin{figure}[t!]
\includegraphics[trim={0cm 0cm 0cm 0cm},clip,scale=.27]{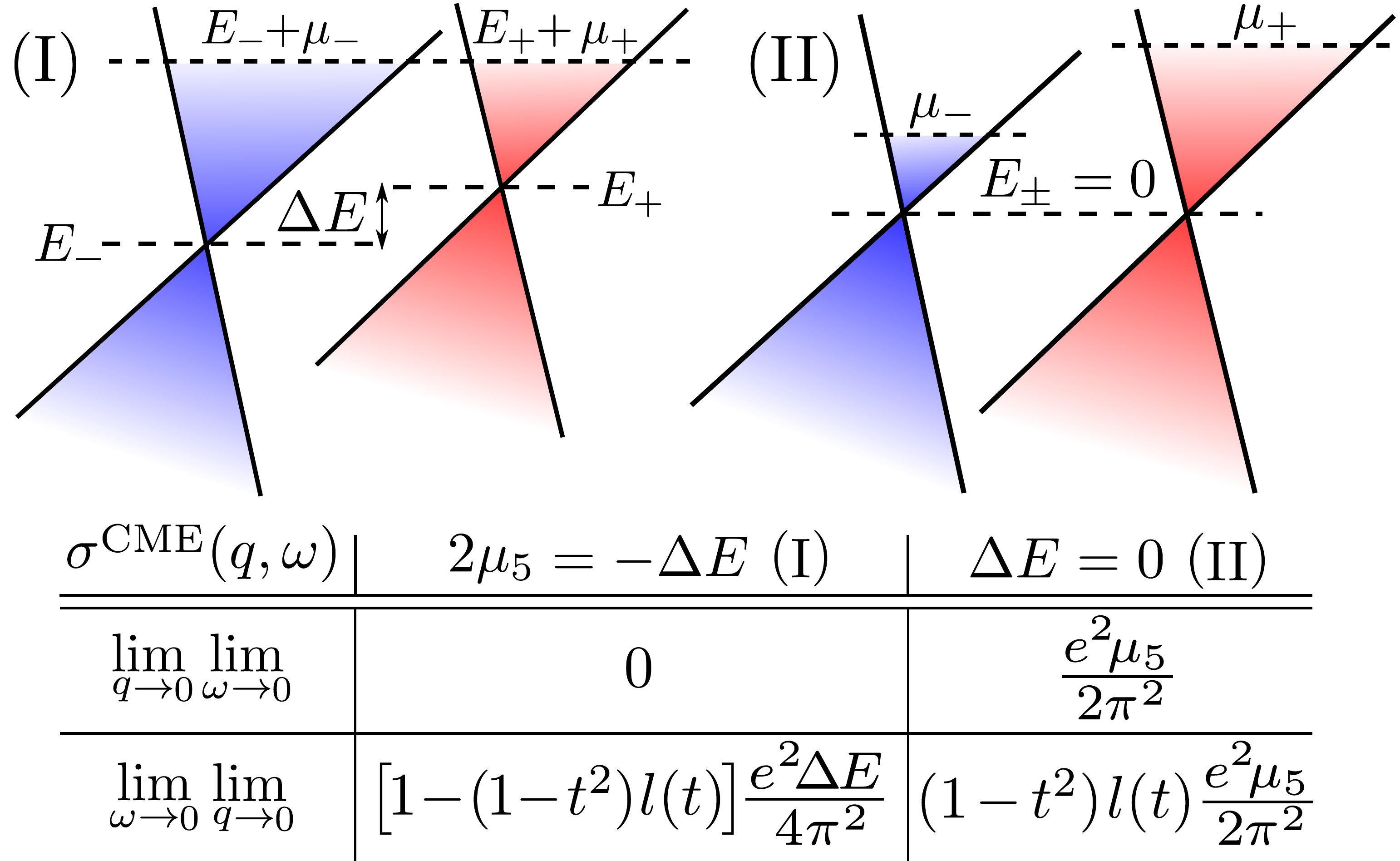}
\caption{Results for the long-wavelength limit of the chiral magnetic conductivity. We show the cases (I) where $2\mu_5 = -\Delta E$ and (II) where the Weyl node separation is zero, i.e., $\Delta E = 0$. The zero-tilt results are\cite{Moore, Pesin1} (bottom, left) $e^2\Delta E/6\pi^2$, and (bottom, right) $e^2\mu_5/6\pi^2$. The response in the second row is sometimes referred to as gyrotropy. \label{fig:table}}
\vspace{-.2cm}
\end{figure}

Here, we reconsider these subtleties for a pair of type-I tilted Weyl cones. We first illustrate the short-wavelength physics involved by calculating the full frequency and wave-number dependence of the effective CME for a transverse electromagnetic wave propagating along the tilting direction. For arbitrary magnetic field directions we focus on the long-wavelength response. We find that the chiral magnetic conductivity is anisotropic and in general nonuniversal, even though the chiral anomaly is unmodified by the tilt. Our results for the homogeneous and static limit are summarized in \figgref{fig:table}.

\textit{Current-current response function.---}
We consider a pair of Weyl cones with opposite chiralites $\pm$ that are doped with chemical potentials $\mu_{\pm} \equiv \mu \pm \mu_5$, defined with respect to the Weyl nodes, as depicted in \figgref{fig:FrontPage}. The chiral chemical potential $\mu_5$ indicates a chiral population imbalance that can be created by applying an electric field pulse with a component parallel to an already present magnetic field. We also allow the Weyl nodes to be split up in energy, which we denote by $\Delta E \equiv E_+ - E_-$, and we comment on the effect of this later on. The topological anomalous Hall effect, however, is well understood and therefore not discussed throughout the following. Furthermore, we consider for simplicity cones with an isotropic Fermi velocity $v_F$, which is straightforwardly generalized to the anisotropic case.

Tilting the cones in a direction $\bt$ can be achieved in two distinct ways. Either we introduce a momentum-dependent chiral chemical potential $\mu_5(\bk) \equiv \mu_5 - \hbar v_F\bk \cdot \bt$, or a momentum-dependent chemical potential $\mu(\bk) \equiv \mu - \hbar v_F\bk\cdot\bt$, where $\hbar\bk$ is the momentum. Only the latter replacement breaks inversion symmetry \cite{Fritz2}. Physically, breaking inversion symmetry corresponds to tilting the two cones in the same direction (c.f.\ \figgref{fig:FrontPage}), while inversion symmetry is preserved upon tilting the two cones in opposite directions. In this paper we perform all calculations explicitly in the case that inversion symmetry is broken, and we comment on the other case in our discussion. Hence, the appropriate Hamiltonian reads ($\hbar=1$)
\be \label{eq:hamil2}
{H}(\bk) = (v_F\bk\cdot{\bm \s} - \mu_5\s^0)\tau^z + (v_F\bk\cdot\bt - \mu)\tau^0\s^0,
\ee
where ${\bm \tau}$ are the Pauli matrices acting in orbital space and ${\bm \s}$ in spin space, complemented by the $2\times2$ unit matrices $\tau^0$ and $\s^0$. The Hamiltonian has four distinct eigenvalues $\s E_{\bk} + v_F\bk\cdot\bt - \mu_{\s'}$, with $E_{\bk} \equiv v_F |\bk|$ the dispersion relation of the massless fermions and $\s,\s' = \pm$. Here, we consider type-I (semi)metals, meaning that we restrict ourselves to $0<t<1$ for $t = |\bt|$. For simplicity we consider the two cones to have the same absolute value for the tilt $t$, but also this is easily generalized.

In order to calculate the response to an externally applied magnetic or electric field, we couple the fermions with charge $-e$ to an external vector potential ${\bf A}$ via the minimal coupling prescription $\bk \rightarrow \bk + e {\bf A}$. Next, we perform second-order perturbation theory in the external gauge field to obtain the current-current response function $\Pi^{ij}(\bq,\w;\bt)$. In the process the subtraction of the two Dirac seas of the cones leads to the elimination of a logarithmic ultra-violet divergence. In terms of the frequency $\w^+  = \w + i0$, the antisymmetric part of the retarded current-current response function $\Pi_l(\bq,\w;\bt) = \eps_{ijl} \Pi^{ij}(\bq,\w;\bt)/2$ reads
\ba \label{eq:pol}
&i\Pi_{l}(\bq,\w;\bt) = \frac{e^2v_F^2}{2}\!\!\!\sum_{\s,\s',\s''}\!\!\s\!\int \!\! \frac{d^3\bk}{(2\pi)^3} F_l^{\s'\!\s''}(\bk,\bq;\bt) \nonumber\\
&\!\!\!\!\times\!\!\bigg[\!\frac{\NFD{E_{\bk}\!-\!\s'\mu_{\s}(\bk)}\!-\!\s'\s''\!\NFD{E_{\bk+\bq}\!-\!\s''\mu_{\s}(\bk\!+\!\bq)}}{\w^+ - v_F\bq\cdot \bt + \s' E_{\bk} - \s''E_{\bk+\bq}}\!\bigg]\!,\!\!\!
\ea
where we defined a structure factor $F_l^{\s\s'}(\bk,\bq;\bt)$ by
\be \label{eq:structure}
{\bf F}^{\s\s'}\!(\bk,\bq;\bt) \equiv\! \frac{\bk}{|\bk|} - \s\s'\frac{\bk\!+\!\bq}{|\bk\!+\!\bq|} - \s'\frac{\bq(\bt\cdot\bk)\! -\! (\bq\cdot\bt)\bk}{|\bk\!+\!\bq||\bk|}.
\ee
In \eqqref{eq:pol} we denoted the Fermi-Dirac distribution by
$\NFD{x} \equiv (e^{x/k_B T} + 1)^{-1}$ and all three sums run over $\s,\s',\s''=\pm$. Physically, the sum over $\s$ accounts for the two cones, whereas the sums over $\s'$ and $\s''$ account for the four possibilities for particle-hole pairs in a chiral cone consisting of two touching bands. In the limit ${\bf t} = {\bf 0}$ the expression in \eqqref{eq:pol} reduces to the well-known result for a three-dimensional chirally doped Weyl semimetal \cite{Warringa}. Including a tilt alters the energy-dispersion relation and yields an additional term in the interaction vertex, resulting in the last term in the structure factor in \eqqref{eq:structure}.

The antisymmetric part of the current-current response function in \eqqref{eq:pol} is in general\cite{Explain1} spanned by a linear combination of the vectors $\bq$ and $\bt$, i.e., we can decompose it as
\be \label{eq:decomp}
i\Pi_l(\bq,\w^+;\bt) = \s^{\text{CME}}(\bq,\w) q_l + \s^{\text{AHE}}(\bq,\w)\w t_l.
\ee
As explicitly indicated this gives rise to two distinct effects: a chiral magnetic effect and a tilt-induced planar intrinsic anomalous Hall effect (AHE) \cite{Pesin2,Carbotte,Tiwari}. The corresponding currents read
\ba
&{\bf J}^{\text{CME}}(\bq,\w) = \s^{\text{CME}}(\bq,\w) {\bf B}(\bq,\w), \label{eq:CME} \\
&{\bf J}^{\text{AHE}}(\bq,\w) = \s^{\text{AHE}}(\bq,\w) \bt \times {\bf E}(\bq,\w), \label{eq:AHE}
\ea
in terms of the chiral magnetic and anomalous Hall conductivities $\s^{\text{CME}}(\bq,\w)$ and $\s^{\text{AHE}}(\bq,\w)$, respectively. The intimate relation between these two effects is even more clear in relativistic notation, where we have that $\Pi^{\kappa\nu}=i\epsilon^{\kappa\lambda\mu\nu}P_\lambda q_\mu$ and thus $J^{\kappa}=\Pi^{\kappa\nu}A_{\nu} = \epsilon^{\kappa\lambda\mu\nu}P_\lambda F_{\mu\nu}/2$, where $F_{\mu\nu}$ is the Faraday tensor and $P^\lambda = (\sigma^{\rm CME}, \sigma^{\rm AHE} \bt)$ elegantly combines the two conductivities. Note that the gauge invariance of the result is then also manifest.

In the following, we discuss the tilt dependence of both effects separately. In principle, both $\sigma^{\text{CME}}(\bq,\w)$ and $\sigma^{\text{AHE}}(\bq,\w)$ depend on the angle between $\bq$ and $\bt$. In order to make analytic progress, however, we specialize to zero temperature and first consider as an illustrative example the propagation of a purely transverse electromagnetic wave (light) with $\bq\parallel\bt$ for arbitrary wavenumbers and frequencies. This case corresponds to ${\bf B} \perp \bt$, ${\bf E} \perp \bt$, and ${\bf B} \perp {\bf E}$, as the magnetic field is given in momentum space by ${\bf B}(\bq,\w) = i \bq \times {\bf A}(\bq,\w) = \bq \times {\bf E}(\bq,\w)/\w$, and gives an effective CME response that, interestingly, is a combination of the chiral magnetic and anomalous Hall effects.

\textit{Effective chiral magnetic effect for a transverse wave with $\bq \parallel \bt$.---}
In the above case the total current along the magnetic field is determined by the effective CME conductivity $\s_{\perp}^{\text{CME}}(\bq,\w) \equiv iq^l \Pi_l(\bq,\w^+;\bt)/q^2 = \s^{\text{CME}}(\bq,\w) + \s^{\text{AHE}}(\bq,\w)\w t/q$, with $q = |\bq|$. The details of the calculation can be found in the Supplemental Material \cite{SupMat}. Ultimately we find for the effective chiral magnetic conductivity
\be \label{eq:cond1}
\s_{\perp}^{\text{CME}}(q,\w) = \frac{e^2}{4\pi^2} \sum_{\s = \pm}\s\mus
\mathcal{S}_{\perp}^{\text{CME}}\bigg(\frac{\w^+}{v_Fq} - t, \frac{\mus}{v_Fq}; t\bigg).
\ee
The dimensionless function $\mathcal{S}_{\perp}^{\text{CME}}(x,y;t)$ captures all frequency, wavenumber and tilt-dependence of the conductivity. It is given by
\be \label{eq:F3}
\mathcal{S}_{\perp}^{\text{CME}}(x,y;t)\!=\!\frac{1-x^2}{2(1+xt)}-\!\!\!\!\sum_{\s,\s'=\pm}\!\!\!\!\!\s K_{\s'}(x,y;t)H_{\s\s'}(x,y;t),
\vspace{-.1cm}
\ee
in terms of the dimensionless functions
\ba
&K_{\s}(x,y;t) \equiv \bigg(\frac{1-x^2}{16y}\bigg)\bigg[\bigg( \frac{2\s  y + t + x}{1 + x t}\bigg)^2 - 1 \bigg], \label{eq:K}\\
&H_{\s\s'}(x,y;t) \equiv \log\bigg(\!1 + \frac{2y}{(1-\s\s' t)(\s' x-\s)}\bigg) \label{eq:H}.
\ea
The expression for the conductivity in \eqqref{eq:cond1} has a nontrivial dependence on the wavenumber $q$ and frequency $\w$ of the externally applied field. In fact, it is a function of the fraction $\w/v_Fq$, giving a different result in the homogeneous limit and the static limit. Indeed, in the static limit $(\w/v_Fq\rightarrow 0)$, we find the well-known\cite{Burkov1,Bernevig} universal result $e^2\mu_5/2\pi^2$,
\begin{figure}[t!]
\includegraphics[trim={0cm 0cm 0cm 0cm},clip,scale=.51]{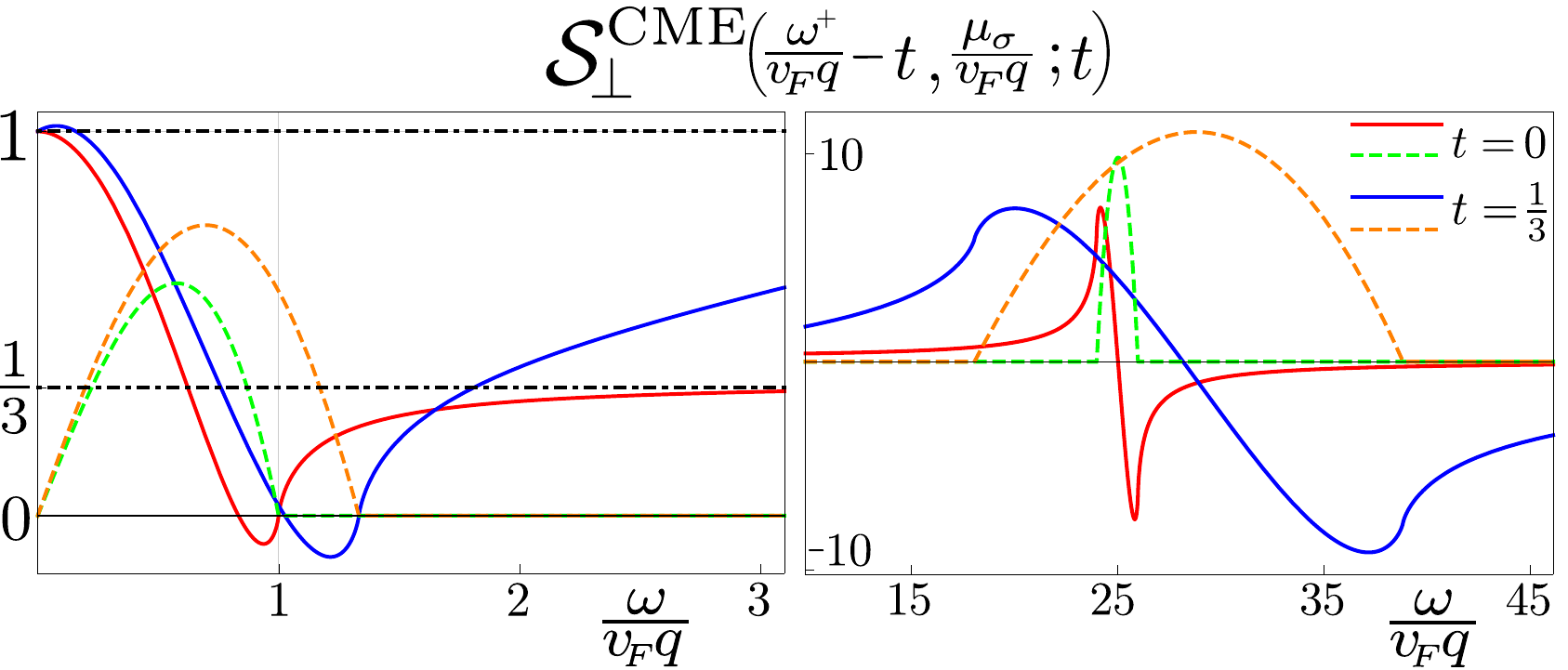}
\caption{Plot of the real (solid lines) and imaginary (dashed lines) part of $\mathcal{S}^{\text{CME}}_\perp$ for $2\mus/v_Fq = 25$ and for $t=0$ (red, green) and $t=1/3$ (blue, orange). The left plot shows the behavior for small $\w/v_Fq$, whereas the right plot shows the resonance for larger values of $\w/v_Fq$. The black dotted-dashed lines indicate the static and homogeneous limit for $t=0$.} \label{fig:CME}
\vspace{-.1cm}
\end{figure}
whereas in the homogeneous limit $(\w/v_Fq \rightarrow \infty)$, we find the tilt-dependent result
\be \label{eq:cond1homg}
\limhomg \s_{\perp}^{\text{CME}}(q,\w) \rightarrow \left[1- 2l(t) + tl(t)\frac{\w }{v_F q}\right]\!\frac{e^2\mu_5}{2\pi^2},
\ee
in terms of the function
\be \label{eq:tfunc}
l(t) \equiv \frac{1}{2t^3}\log \left( \frac{1+t}{1-t} \right)-\frac{1}{t^2} \text{ } \overset{t\rightarrow0}{=} \text{ } \frac{1}{3}.
\ee
We thus obtain the result $e^2\mu_5/6\pi^2$ for the homogeneous limit of \eqqref{eq:cond1homg} if $t=0$ \cite{Pesin1}. The function $l(t)$ diverges upon taking the limit $t\rightarrow1$. The physical reason for this divergence is that then the cones are tilted up to the point that the density of states becomes infinite, thus resulting in an infinite conductivity. In fact, the conductivity in \eqqref{eq:cond1homg} is due to the presence of the in-plane anomalous Hall effect
formally always infinite in the homogeneous limit $\w/v_F q \rightarrow\infty$. Note, however, that for light propagation we have that $\w/q$ is equal to the speed of light in the material.

We plot the full dependence of the real and imaginary part of $\mathcal{S}_{\perp}^{\text{CME}}\big(\w^+\!/v_Fq-t, \mus\!/v_Fq, t\big)$ on $\w/v_Fq$ for a fixed value of $\mus/v_Fq$ and different values of the tilt $t$ in \figgref{fig:CME}. When $t=0$, the real part interpolates between the value $1$ in the static limit and $1/3$ in the homogeneous limit \cite{Warringa,Sau}. Additionally, there is a resonance at $\w = 2\mus$, after which the conductivity goes to zero as $1/\w$. This resonance is effectively shifted to infinity in the homogeneous limit. For a nonzero tilt $t$, the static limit remains unchanged and in the homogeneous limit the real part of the conductivity diverges as $tl(t) \w/v_Fq$, rather than becoming constant as in the case of zero tilt. The resonance at $\w = 2\mus$ remains present at nonzero tilt but becomes broader, as its width is now set by $2\mus/(1\pm t)$. Note that the conductivity $\s_{\perp}^{\text{CME}}(q,\w)$ depends in a highly nonlinear way on the chiral imbalance $\mu_5$. Theoretically this implies that the CME is not fully determined by the triangle diagram of the chiral anomaly. This is only true in the long-wavelength limit \cite{Jackiw2,Victoria}.
\begin{figure}[t!]
\includegraphics[trim={0cm 0cm 0cm 0cm},clip,scale=.6]{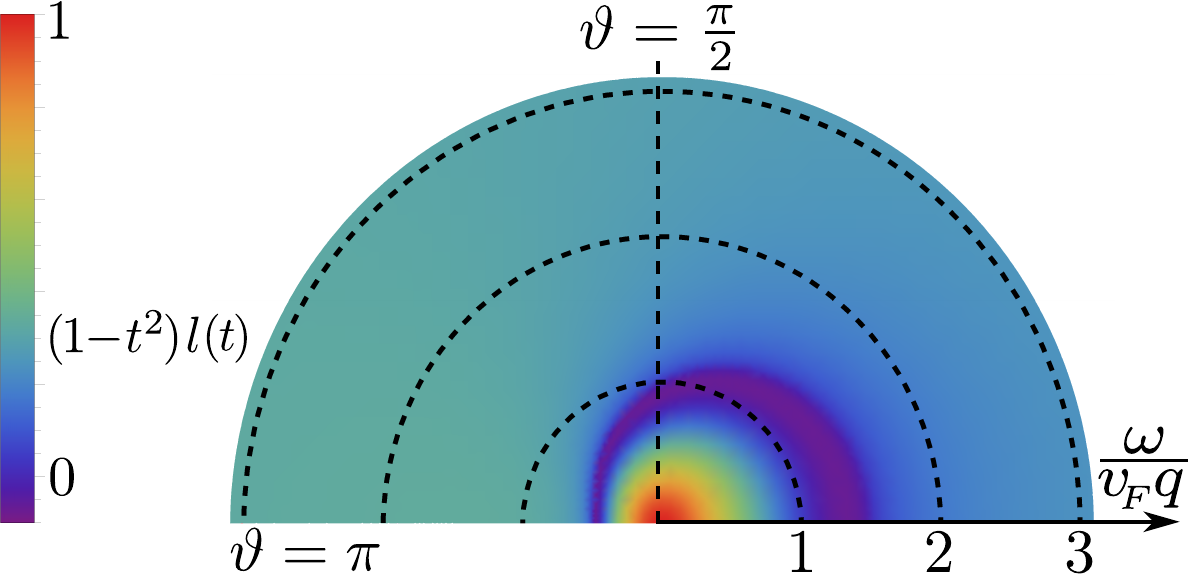}
\caption{Anisotropic behavior of $\text{Re}[\mathcal{S}^{\text{CME}}]$ for $t = 1/2$ as a function of the angle $\vartheta$ and the radial coordinate $\w/v_F q$. In the static (small radius) and homogeneous (large radius) limit, we obtain the isotropic results $1$ and $(1-t^2)l(t) \simeq 0.3$. \label{fig:angle}}
\vspace{-.2cm}
\end{figure}

At this point it is important to discuss why the conductivity is finite in the static limit in equilibrium. In deriving \eqqref{eq:pol} a logarithmic divergence was avoided by a cancellation of the Dirac-sea contributions of the two cones. This cancellation is correct up to a constant, which is proportional to the energy separation $\Delta E = E_+ - E_-$ of the Weyl nodes \cite{Kharzeev1,Kharzeev2,Landsteiner}. Hence, our answers for the static limit and \eqqref{eq:cond1homg} only apply when the energy separation between the nodes is zero. If that is not the case, then the true equilibrium situation corresponds to the situation where the chiral imbalance is exactly canceled by the energy separation between the Weyl nodes, i.e., $2\mu_5 = \mu_+ - \mu_- = -\Delta E$. Using this renormalization condition, we find that the chiral magnetic conductivity is zero in equilibrium, as expected. We will follow the same procedure when we consider a general angle between the externally applied magnetic field and the tilt direction.

\textit{Angle dependence of the chiral magnetic effect.---}
We define $\vartheta$ to be the angle between ${\bf q}$ and $\bt$, such that ${\bf q} \cdot\bt = q t \cos\vartheta \equiv qt_\parallel$. For arbitrary angles $\vartheta$ we cannot perform the necessary integrals analytically for all wavenumbers $q$ and frequencies $\w$. However, we can investigate the tilt dependence of the long-wavelength limit of the conductivity for arbitrary angles. To do so, we take the limit $q\rightarrow0$ in the integrand of \eqqref{eq:pol}, while keeping $\w/v_F q$ fixed. Keeping in mind that we are not considering a possible topological contribution to the anomalous Hall effect, we find in general the interesting relation $\s^{\text{AHE}}(\bq,\w) = \s^{\text{CME}}(\bq,\w)/v_F(1-t^2)$ with
\be \label{eq:cond1}
\s^{\text{CME}}(\bq,\w) = \frac{e^2\mu_5}{2\pi^2} \mathcal{S}^{\text{CME}}\bigg(\frac{\w^+}{v_Fq} - t_\parallel; \bt\bigg).
\ee
The dimensionless function $\mathcal{S}^{\text{CME}}(x;\bt)$ is given by
\be
\!\!\mathcal{S}^{\text{CME}}(x;\bt) = \frac{(1-t^2)(1-x^2)}{N^2(x,\bt)}\bigg[1\! +\! \frac{(x\!+\!\tpara)H(x;\bt)}{2N(x;\bt)}\bigg],
\ee
in terms of $N(x;\bt) \equiv \sqrt{(1 - t^2)(1-x^2)+(x+\tpara)^2}$, $H(x;\bt) \equiv \sum_{\s=\pm}\s\log[(x-\s)M_{\s}(x;\bt)]$,
and finally also\cite{SupMat} $M_{\s}(x;\bt) \equiv (1 + N(x;\bt) + \tpara x)(1-\s \tpara) - (t^2-\tpara^2) (1+\s x)$. In \figgref{fig:angle} we show the resulting angular dependence of the conductivities. In the static limit we have that the chiral magnetic conductivity is equal to $e^2\mu_5/2\pi^2$ and is independent of the tilt \cite{Bernevig}. In the homogeneous limit we find the result $(1-t^2) l(t) e^2\mu_5/2\pi^2$  for all angles \cite{Explain2}. Again, we need to add an appropriate renormalization constant for $\Delta E \neq 0$ such that the chiral magnetic current is zero in equilibrium. Using this subtraction procedure, which in particle physics amounts to adding a Bardeen counterterm \cite{Landsteiner}, we find a general answer that depends on $\mu_5$ and $\Delta E$ and modifies the results in the homogeneous limit. We displayed the final results for the special cases of equilibrium and zero energy separation between the cones in \figgref{fig:table}.
\begin{figure}[t!]
\includegraphics[trim={0cm 0cm 0cm 0cm},clip,scale=.45]{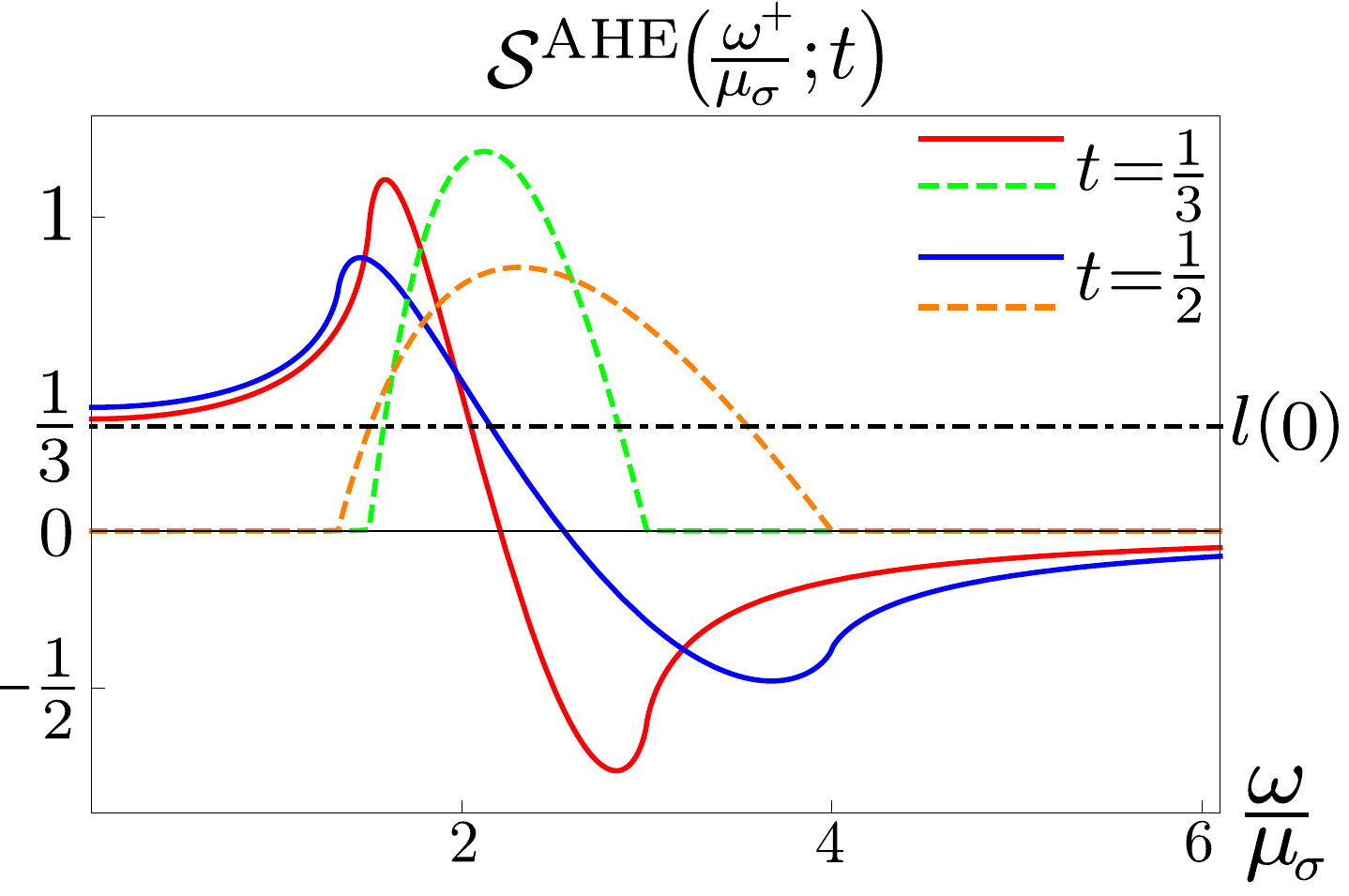}
\vspace{-.3cm} 
\caption{Plot of the real (solid lines) and imaginary (dashed lines) parts of the function $\mathcal{S}^{\text{AHE}}(\omega^+/\mus;t)$ for $t=1/3$ (red, green), and $t=1/2$ (blue, orange) \cite{Pesin2}. The black dotted-dashed line indicates the tilt-independent limit.} \label{fig:AHE}
\vspace{-.2cm}
\end{figure}

\textit{Frequency dependence of the anomalous Hall effect.---}
As advertized the tilt induces another interesting effect, namely a planar intrinsic anomalous Hall effect with a current given by \eqqref{eq:AHE} that is perpendicular to both the external electric field and the tilting direction. Apart from the long-wavelength limit following from \eqqref{eq:cond1}, we are also able to obtain the full frequency dependence of the homogeneous anomalous Hall conductivity as \cite{SupMat}
\be
\s^{\text{AHE}}({\bf 0}, \w) = \frac{e^2}{4\pi^2v_F}\sum_{\s = \pm} \s \mus \mathcal{S}^{\text{AHE}}\Big(\frac{\w^+}{\mus};t\Big),
\ee
in terms of the dimensionless function
\be
\mathcal{S}^{\text{AHE}}(y;t) = -\frac{1}{2t^2} + \sum_{\s,\s'}\s'L_{\s}(y;t)H_{\s\s'}(0,1/y;t),
\ee
where $L_{\s}(y;t)\equiv \big[y^2t^2-(2-\s y)^2\big]/16yt^3$ and again we encounter the functions $H_{\s\s'}(x,y;t)$ from \eqqref{eq:H}. This result was recently obtained in a different way both analytically \cite{Pesin2,Carbotte} and numerically \cite{Tiwari}. In the zero-frequency limit the conductivity reduces to
$\s^{\text{AHE}}({\bf 0}, 0) =  l(t) e^2\mu_5/2\pi^2v_F$,
which corresponds exactly to the slope of the linear divergence in \eqqref{eq:cond1homg}. We plot the dependence of the real and imaginary part of $\mathcal{S}^{\text{AHE}}\big(\w^+/\mus;t)$ on $\w/\mus$ in \figgref{fig:AHE} for several magnitudes of the tilt. Again, we observe a resonance behavior around $\w=2\mus$, similar to the one in \figgref{fig:CME}, because the current response in that figure is dominated by the AHE at large frequencies.

\textit{Discussion.---}
We have shown that the electric and magnetic response of a pair of tilted Weyl cones is in general non-universal and depends on the magnitude of the tilt and on the angle between the tilt direction and the wave vector of the magnetic field. However, the chiral anomaly is due to the lowest Landau level, which only obtains a change of slope due to a tilting of the cones \cite{Ninomiya}. Hence, we expect the chiral anomaly to be unmodified and thus isotropic. Using the relation between the current-current correlation function and the triangle diagram in the static (adiabatic) limit, we find for the time derivative of the chiral number density $n_5 \equiv n_+ - n_-$,
\be
\frac{d n_5}{dt} = \lim_{\w\rightarrow0}{\frac{e^2}{2\pi^2}} \mathcal{S}^{\text{CME}}\bigg(\frac{\w^+}{v_Fq} - t_\parallel; \bt \bigg){\bf E}\cdot {\bf B}.
\ee
In the static limit we have $\mathcal{S}^{\text{CME}}(\w^+/v_Fq - t_\parallel; \bt ) \rightarrow 1$, such that we indeed find an unmodified chiral anomaly.

Additionally, we showed that the chiral magnetic effect is closely related to an in-plane tilt-induced anomalous Hall effect, for which we calculated the dynamical conductivity. We have also performed all these calculations in the case that inversion symmetry is not broken, corresponding to tilting the Weyl cones in opposite directions. An important consequence is that in the long-wavelength limit the anomalous Hall effect becomes proportional to $2\mu$, instead of $2\mu_5$, i.e., $\s^{\text{AHE}}({\bf 0}, 0) =  l(t)e^2\mu/2\pi^2v_F$. The chiral magnetic effect, however, remains proportional to $2\mu_5$ due to Bloch's theorem \cite{Yamamoto}.

It is our pleasure to thank Guido van Miert for useful discussions and a critical reading of the manuscript. This work is supported by the Stichting voor Fundamenteel Onderzoek der Materie (FOM) and is part of the D-ITP consortium, a program of the Netherlands Organisation for Scientific Research (NWO) that is funded by the Dutch Ministry of Education, Culture and Science (OCW).

\onecolumngrid


\section*{Supplementary material}

# Supplemental Material for "Anisotropic chiral magnetic effect from tilted Weyl cones"

E.C.I. van der Wurff* and H. T. C. Stoof

*Institute for Theoretical Physics and Center for Extreme Matter and Emergent Phenomena,*
*Utrecht University, Princetonplein 5, 3584 CC Utrecht, The Netherlands*

In this supplemental material we provide detailed derivations of some of the results given in the main text. We start by explicitly deriving the form of the antisymmetric part of the current-current response function. Subsequently, we calculate the full wavenumber and frequency-dependence of the effective chiral magnetic conductivity for a transverse wave. Furthermore, we calculate the long-wavelength limit of the chiral magnetic conductivity for arbitrary angles between the tilting direction and the wavenumber of the external field. Finally, we show the calculation of the full-frequency behavior of the planar intrinsic anomalous Hall conductivity.

## CONTENTS

## I. DERIVATION ANTISYMMETRIC PART CURRENT-CURRENT RESPONSE FUNCTION

In this section we derive the antisymmetric part of the current-current response function as given in Eq. (2) in the main text. We describe the two Weyl fermions with opposite chirality by the four-dimensional Dirac spinor $\psi$ and denote the corresponding chiral imbalance by $\mu_5$. Setting $v_F = \hbar = 1$ for the moment, the Hamiltonian in Eq. (1) in the main text corresponds to the following quadratic action in Fourier space

$$S[\bar\psi, \psi] = \int \frac{d^4k}{(2\pi)^4} \bar\psi(k) \big[ \slashed{k} - \mu\gamma^0 - \mu_5 \gamma^0 \gamma^5 + (\mathbf{k}\cdot\mathbf{t})\gamma^0 \big] \psi(k), \qquad (1)$$

where $\gamma^\mu$ are the Dirac gamma matrices and $\slashed{k} = k_\mu \gamma^\mu = \eta_{\mu\nu} k^\nu \gamma^\mu$ in terms of the Minkowski metric $\eta_{\mu\nu} = \text{diag}(-,+,+,+)$ and the four-momentum $k^\mu = (\omega, \mathbf{k})$. The corresponding inverse fermion propagator reads $S_F^{-1}(k) = -i\big[\slashed{k} - \mu\gamma^0 - \mu_5 \gamma^0 \gamma^5 + (\mathbf{k}\cdot\mathbf{t})\gamma^0\big]$. Our conventions for the gamma matrices $\gamma^\mu$ can be found in Appendix A. We proceed by coupling the Weyl fermions with charge $-e$ to an external gauge field $A_\mu(k)$ via the minimal coupling prescription. This yields a coupling term in the action of the form

$$J^\mu(k) A_\mu(-k) = \big[e\bar\psi(k)\gamma^\mu \psi(k)\big] A_\mu(-k) + \big[e\bar\psi(k)\gamma^0 t^i \psi(k)\big] A_i(-k). \qquad (2)$$

Subsequently, we perform second-order perturbation theory in the coupling to find that the effective action for the gauge field reads

$$S^{\text{eff}}[A] = -\frac{1}{2} \int \frac{d^4q}{(2\pi)^4} A_\mu(q) \Pi^{\mu\nu}(q) A_\nu(-q), \qquad (3)$$

---

* e.c.i.vanderwurff@uu.nl

with $\Pi^{\mu\nu}(q)$ the polarization tensor of the Weyl (semi)metal. Focusing on the spatial components, it reads

$$\Pi^{ij}(q) = ie^2 \int \frac{d^4k}{(2\pi)^4} \text{Tr}\Big[S_F(k+q)(\gamma^i + t^i\gamma^0)S_F(k)(\gamma^j + t^j\gamma^0)\Big]. \quad (4)$$

To proceed, it is useful to define

$$iG_F^{-1}(k) \equiv i\gamma^0 S_F^{-1}(k) = \begin{pmatrix} \omega + \mathbf{k}\cdot\boldsymbol{\sigma} + \mu_-(\mathbf{k}) & 0 \\ 0 & \omega - \mathbf{k}\cdot\boldsymbol{\sigma} + \mu_+(\mathbf{k}) \end{pmatrix} \equiv \begin{pmatrix} G_-^{-1}(k) & 0 \\ 0 & G_+^{-1}(k) \end{pmatrix}, \quad (5)$$

where we also defined the momentum-dependent chemical potentials $\mu_\pm(\mathbf{k}) \equiv \mu \pm \mu_5 - \mathbf{k}\cdot\mathbf{t}$. Inverting the Green's function yields $G_\pm(k) = \mathcal{G}_\mu^\pm(k)\sigma^\mu$, with $\sigma^\mu = (1,\boldsymbol{\sigma})$ and

$$\mathcal{G}_0^\pm(k) \equiv \frac{\omega + \mu_\pm(\mathbf{k})}{(\omega + \mu_\pm(\mathbf{k}))^2 - \mathbf{k}^2} \quad \text{and} \quad \mathcal{G}_i^\pm(k) \equiv \frac{\pm k_i}{(\omega + \mu_\pm(\mathbf{k}))^2 - \mathbf{k}^2}. \quad (6)$$

Using $S_F(k) = G_F(k)\gamma^0$ we rewrite the polarizability as

$$\Pi^{ij}(q) = -ie^2 \int \frac{d^4k}{(2\pi)^4} \Big\{ \mathcal{G}_\alpha^-(k+q)\mathcal{G}_\beta^-(k)\text{Tr}\big[\sigma^\alpha(\sigma^i - t^i)\sigma^\beta(\sigma^j - t^j)\big] + \mathcal{G}_\alpha^+(k+q)\mathcal{G}_\beta^+(k)\text{Tr}\big[\sigma^\alpha(\sigma^i + t^i)\sigma^\beta(\sigma^j + t^j)\big] \Big\}. \quad (7)$$

We extract the antisymmetric part by writing $\Pi_l(q) = \epsilon_{ijl}\Pi^{ij}(q)/2$. For the trace we find

$$\epsilon_{ijl}\text{Tr}\big[\sigma^\alpha(\sigma^i \pm t^i)\sigma^\beta(\sigma^j \pm t^j)\big] = 4i\Big(\delta_l^\alpha\delta^{\beta 0} - \delta^{\alpha 0}\delta_l^\beta \mp t^\alpha\delta_l^\beta \pm t^\beta\delta_l^\alpha\Big). \quad (8)$$

For briefness, we write $\Pi_l(q) = \Pi_l^{(1)}(q) + \Pi^{(2)}(q)$, with

$$\Pi_l^{(1)}(q) = 2e^2 \sum_\sigma \int \frac{d^4k}{(2\pi)^4} \Big[\mathcal{G}_l^\sigma(k+q)\mathcal{G}_0^\sigma(k) - \mathcal{G}_0^\sigma(k+q)\mathcal{G}_l^\sigma(k)\Big],$$

$$\Pi_l^{(2)}(q) = 2e^2 \sum_\sigma \sigma \int \frac{d^4k}{(2\pi)^4} \Big[\mathcal{G}_l^\sigma(k+q)\mathcal{G}_i^\sigma(k)t^i - \mathcal{G}_i^\sigma(k+q)t^i\mathcal{G}_l^\sigma(k)\Big]. \quad (9)$$

Here $\Pi^{(2)}(q)$ includes the effects of the corrections on the current operator due to the tilt of the Weyl cones. We start by considering $\Pi^{(1)}(q)$. Upon Wick rotating, we obtain in terms of the complex frequency $z$ and the fermionic Matsubara frequency $i\omega_n$,

$$\Pi_l^{(1)}(\mathbf{q}, z; \mathbf{t}) = \frac{2ie^2}{\beta} \sum_{\sigma, i\omega_n} \int \frac{d^3k}{(2\pi)^3} \Big[\mathcal{G}_l^\sigma(\mathbf{k}+\mathbf{q}, i\omega_n + z)\mathcal{G}_0^\sigma(\mathbf{k}, i\omega_n) - \mathcal{G}_0^\sigma(\mathbf{k}+\mathbf{q}, i\omega_n + z)\mathcal{G}_l^\sigma(\mathbf{k}, i\omega_n)\Big]$$

$$= \frac{ie^2}{2}\sum_\sigma \sigma \int \frac{d^3k}{(2\pi)^3}\bigg[N_{\mathrm{F}}(|\mathbf{k}| - \mu_\sigma(\mathbf{k}))\bigg(-\frac{F_l^-(\mathbf{k},\mathbf{q})}{z - \mathbf{q}\cdot\mathbf{t} + |\mathbf{k}| - |\mathbf{k}+\mathbf{q}|} - \frac{F_l^+(\mathbf{k},\mathbf{q})}{z - \mathbf{q}\cdot\mathbf{t} + |\mathbf{k}| + |\mathbf{k}+\mathbf{q}|}\bigg)$$

$$+ N_{\mathrm{F}}(-|\mathbf{k}| - \mu_\sigma(\mathbf{k}))\bigg(\frac{F_l^+(\mathbf{k},\mathbf{q})}{z - \mathbf{q}\cdot\mathbf{t} - |\mathbf{k}| - |\mathbf{k}+\mathbf{q}|} + \frac{F_l^-(\mathbf{k},\mathbf{q})}{z - \mathbf{q}\cdot\mathbf{t} - |\mathbf{k}| + |\mathbf{k}+\mathbf{q}|}\bigg)$$

$$+ N_{\mathrm{F}}(|\mathbf{k}+\mathbf{q}| - \mu_\sigma(\mathbf{k}+\mathbf{q}))\bigg(\frac{F_l^-(\mathbf{k},\mathbf{q})}{z - \mathbf{q}\cdot\mathbf{t} + |\mathbf{k}| - |\mathbf{k}+\mathbf{q}|} - \frac{F_l^+(\mathbf{k},\mathbf{q})}{z - \mathbf{q}\cdot\mathbf{t} - |\mathbf{k}| - |\mathbf{k}+\mathbf{q}|}\bigg)$$

$$+ N_{\mathrm{F}}(-|\mathbf{k}+\mathbf{q}| - \mu_\sigma(\mathbf{k}+\mathbf{q}))\bigg(\frac{F_l^+(\mathbf{k},\mathbf{q})}{z - \mathbf{q}\cdot\mathbf{t} + |\mathbf{k}| + |\mathbf{k}+\mathbf{q}|} - \frac{F_l^-(\mathbf{k},\mathbf{q})}{z - \mathbf{q}\cdot\mathbf{t} - |\mathbf{k}| + |\mathbf{k}+\mathbf{q}|}\bigg)\bigg],$$

where we performed the sum over the fermionic Matsubara frequencies yielding the Fermi-Dirac distributions $N_{\mathrm{F}}(x) \equiv (e^{\beta x} + 1)^{-1}$, with $\beta \equiv 1/k_B T$, and we abbreviated

$$F_l^\pm(\mathbf{k}, \mathbf{q}) \equiv \frac{k_l}{|\mathbf{k}|} \pm \frac{k_l + q_l}{|\mathbf{k}+\mathbf{q}|}. \quad (10)$$

Next, we write for the second and fourth group of terms by using $N_{\mathrm{F}}(-x) = 1 - N_{\mathrm{F}}(x)$. Performing this replacement, we see that the $+1$-term directly drops out for the terms proportional to $F_l^-(\mathbf{k},\mathbf{q})$. The other two terms drop out

individually after summing over $\sigma$. This is the point where we substract two linearly divergent integrals to yield zero, which is correct up to a constant proportional to the energy separation of the Weyl cones $\Delta E$. We use the undetermined proportionality constant to fix the current to be zero in equilibrium. Doing a similar calculation for $\Pi^{(2)}(q)$ we find

$$\Pi_l^{(2)}(\mathbf{q}, z; \mathbf{t}) = \frac{ie^2}{2} \sum_{\sigma,\sigma',\sigma''} \sigma \int \frac{d^3\mathbf{k}}{(2\pi)^3} \left[ \sigma'' \frac{q_l(\mathbf{t} \cdot \mathbf{k}) - (\mathbf{q} \cdot \mathbf{t})k_l}{|\mathbf{k}||\mathbf{k}+\mathbf{q}|} \right] \left[ \frac{N_F(|\mathbf{k}| - \sigma'\mu_\sigma(\mathbf{k})) - \sigma'\sigma'' N_F(|\mathbf{k}+\mathbf{q}| - \sigma''\mu_\sigma(\mathbf{k}+\mathbf{q}))}{z - \mathbf{q}\cdot\mathbf{t} + \sigma' k - \sigma''|\mathbf{k}+\mathbf{q}|} \right], \quad (11)$$

with $\sigma, \sigma', \sigma'' = \pm$. Adding up both contributions and reinstating all factors $v_F$ we arrive at

$$\boxed{\begin{aligned} i\Pi_l(\mathbf{q}, z; \mathbf{t}) = &\frac{e^2 v_F^2}{2} \sum_{\sigma,\sigma',\sigma''} \sigma \int \frac{d^3\mathbf{k}}{(2\pi)^3} \left[ \frac{k_l}{|\mathbf{k}|} - \sigma'\sigma'' \frac{k_l + q_l}{|\mathbf{k}+\mathbf{q}|} - \sigma'' \frac{q_l(\mathbf{t}\cdot\mathbf{k}) - (\mathbf{q}\cdot\mathbf{t})k_l}{|\mathbf{k}||\mathbf{k}+\mathbf{q}|} \right] \\ &\times \left[ \frac{N_F(E_\mathbf{k} - \sigma'\mu_\sigma(\mathbf{k})) - \sigma'\sigma'' N_F(E_{\mathbf{k}+\mathbf{q}} - \sigma''\mu_\sigma(\mathbf{k}+\mathbf{q}))}{z - v_F \mathbf{q}\cdot\mathbf{t} + \sigma' E_\mathbf{k} - \sigma'' E_{\mathbf{k}+\mathbf{q}}} \right], \end{aligned}} \quad (12)$$

with $E_\mathbf{k} = v_F|\mathbf{k}|$ the dispersion relation of the massless fermions. The terms between the first pair of square brackets define the form factor $\mathbf{F}^{\sigma'\sigma''}(\mathbf{k}, \mathbf{q}; \mathbf{t})$ as used in the main text. Furthermore, note that all the calculations we present in the main text are at zero temperature. This is achieved by the replacement $N_F(x) \to \vartheta(-x)$, in terms of the Heaviside function $\vartheta(x)$.

## II. DECOMPOSITION ANTISYMMETRIC PART CURRENT-CURRENT RESPONSE FUNCTION

Note that with the definitions Eq. (3) and Eq. (4) the current reads

$$J^i(q) = \frac{\delta S^{\mathrm{eff}}[A]}{\delta A_i(-q)} = -\frac{1}{2}\left[\Pi^{ij}(-q) + \Pi^{ji}(q)\right]A_j(q) = \epsilon^{ijl}\Pi_l(q)A_j(q), \quad (13)$$

where in the last equality we only considered the antisymmetric contribution to the current. Hence, if we use the decomposition as presented in the main text, i.e.,

$$i\Pi_l(\mathbf{q}, \omega^+; \mathbf{t}) = \sigma^{\mathrm{CME}}(\mathbf{q}, \omega) q_l + \sigma^{\mathrm{AHE}}(\mathbf{q}, \omega) \omega t_l, \quad (14)$$

we obtain the chiral magnetic and anomalous Hall contributions to the current:

$$\mathbf{J}(\mathbf{q}, \omega) = \sigma^{\mathrm{CME}}(\mathbf{q}, \omega)\mathbf{B}(\mathbf{q}, \omega) + \sigma^{\mathrm{AHE}}(\mathbf{q}, \omega)\mathbf{t} \times \mathbf{E}(\mathbf{q}, \omega). \quad (15)$$

Finally, we can obtain the individual conductivities from the decomposition in Eq. (14) by forming the following linear combinations of the projections $q_l\Pi^l(\mathbf{q}, \omega^+; \mathbf{t})$ and $t_l\Pi^l(\mathbf{q}, \omega^+; \mathbf{t})$:

$$\sigma^{\mathrm{CME}}(\mathbf{q}, \omega) = \frac{t^2(iq_l\Pi^l) - (\mathbf{q}\cdot\mathbf{t})(it_l\Pi^l)}{q^2 t^2 - (\mathbf{q}\cdot\mathbf{t})^2} \quad \text{and} \quad \sigma^{\mathrm{AHE}}(\mathbf{q}, \omega) = \frac{1}{\omega}\frac{q^2(it_l\Pi^l) - (\mathbf{q}\cdot\mathbf{t})(iq_l\Pi^l)}{q^2 t^2 - (\mathbf{q}\cdot\mathbf{t})^2}. \quad (16)$$

Physically there is no pole at $q^2 t^2 - (\mathbf{q}\cdot\mathbf{t})^2 = 0$. Hence, we expect the denominator to exactly drop out once we form the linear combinations in the numerator to find the conductivities.

## III. EFFECTIVE CHIRAL MAGNETIC EFFECT FOR A TRANSVERSE WAVE WITH $\mathbf{q} \parallel \mathbf{t}$

We now consider the case where the external wavenumber of the gauge field is parallel to the tilting direction: $\mathbf{q} \parallel \mathbf{t}$. This corresponds to $\mathbf{B} \perp \mathbf{t}$ as $\mathbf{B}(\mathbf{q}, \omega) = i\mathbf{q} \times A(\mathbf{q}, \omega)$. If we additionally impose that the external gauge field concerns a purely transverse electromagnetic wave, then we also have $\mathbf{E} \perp \mathbf{t}$ and $\mathbf{E} \perp \mathbf{B}$, as $\mathbf{B}(\mathbf{q}, \omega) = \mathbf{q} \times \mathbf{E}(\mathbf{q}, \omega)/\omega$. The effective chiral magnetic conductivity $\sigma^{\mathrm{CME}}_\perp(\mathbf{q}, \omega)$ is then a combination of the chiral magnetic and anomalous Hall effects. It can be found by projecting Eq. (14) onto $q_l$, i.e.,

$$\sigma^{\mathrm{CME}}_\perp(\mathbf{q}, \omega) = \frac{iq_l \Pi^l_\perp(\mathbf{q}, \omega^+; \mathbf{t})}{q^2} = \sigma^{\mathrm{CME}}(\mathbf{q}, \omega) + \sigma^{\mathrm{AHE}}(\mathbf{q}, \omega)\frac{\omega t}{q}, \quad (17)$$

where the subscript "⊥" indicates that this equality is only valid in the case $\mathbf{B} \perp \mathbf{t}$, or $\mathbf{q} \parallel \mathbf{t}$. We now proceed by calculating this effective chiral magnetic conductivity for all frequencies and wavenumbers. Thus, we need to perform the three-dimensional integral after projecting Eq. (12) onto $q_l$. We can put $\mathbf{q}$ along the $z$-axis without loss of generality and subsequently go to spherical coordinates $(\phi, \theta, k)$. Then we have $\mathbf{q} \cdot \mathbf{k} = qk \cos\theta$ and also $\mathbf{k} \cdot \mathbf{t} = kt \cos\theta$, because we are considering the case $\mathbf{q} \parallel \mathbf{t}$. The integral over the azimuthal angle $\phi$ is trivial and yields $2\pi$. The integral over the polar angle can be simplified by two distintinct changes of variables. For the terms proportional to $\vartheta(\pm\mu_\sigma(\mathbf{k}) - E_\mathbf{k})$ we introduce $k' \equiv |\mathbf{k} + \mathbf{q}|$ such that $\mathbf{k} \cdot \mathbf{q} = (k'^2 - k^2 - q^2)/2$. On the other hand, for all terms proportional to $\vartheta(\pm\mu_\sigma(\mathbf{k}+\mathbf{q}) - E_{\mathbf{k}+\mathbf{q}})$ we first perform a shift $\mathbf{k} \to \mathbf{k} - \mathbf{q}$ and subsequently perform a similar change of variables. In the end we find, taking $v_F = 1$ for now,

$$iq_l \Pi_\perp^l(q, z; t) = \frac{e^2}{16\pi^2 q} \sum_{\sigma, \sigma'} \sigma\sigma' \sum_{\sigma', \sigma''} I(q, z; \sigma'\mu_\sigma, t), \tag{18}$$

where

$$I(q, z; \mu_\sigma, t) = \sum_{\sigma', \sigma''} \int_0^\infty dk \int_{|k-q|}^{k+q} dk' \left[\frac{\sigma' f_{\sigma'\sigma''}(k', k, q)}{z' + \sigma'k - \sigma''k'}\right] \vartheta\left(\sigma'\mu_\sigma + \sigma'\frac{tk^2}{2q} + \sigma'\frac{qt}{2} - k - \sigma'\frac{tk'^2}{2q}\right), \tag{19}$$

in terms of $z' \equiv z - \mathbf{q} \cdot \mathbf{t}$ and the function

$$f_{\sigma'\sigma''}(k', k, q) \equiv \sigma'k'(k'^2 - k^2 - q^2) - \sigma''k(k'^2 - k^2 + q^2). \tag{20}$$

After carefully distinguishing the cases when the Heaviside function in Eq. (19) is nonzero, we find

$$I(q, z; \mu_\sigma, t) = \underbrace{\int_0^{\mu_\sigma^+} dk \int_{|k-q|}^{k+q} dk' \left[\frac{f_{++}(k', k, q)}{z' + k - k'} + \frac{f_{+-}(k', k, q)}{z' + k + k'}\right]}_{\equiv I_1(q, z; \mu_\sigma, t)} + \underbrace{\int_{\mu_\sigma^+}^{\mu_\sigma^-} dk \int_{|k-q|}^{\sqrt{P}} dk' \left[\frac{f_{++}(k', k, q)}{z' + k - k'} + \frac{f_{+-}(k', k, q)}{z' + k + k'}\right]}_{\equiv I_2(q, z; \mu_\sigma, t)}, \tag{21}$$

where $P = 2q(\mu_\sigma + tk^2/2q + qt/2 - k)/t$ and $\mu_\sigma^\pm \equiv \mu_\sigma/(1 \pm t)$. The expression $I_2(q, z; \mu_\sigma, t)$ vanishes as $t \to 0$, whereas $I_1(q, z; \mu_\sigma, t)$ reduces to the tilt-independent result when $t \to 0$. To evaluate Eq. (18) we need to calculate $I_1(q, z; \mu_\sigma, t) - I_1(q, z; -\mu_\sigma, t)$, finding

$$I_1(q, z; \mu_\sigma, t) - I_1(q, z; -\mu_\sigma, t) = 2q\mu_\sigma^+(q^2 - z'^2)\left[1 - \sum_{\sigma', \sigma''} F(\sigma'q, \sigma''z', \mu_\sigma^+) \log\left(\frac{2\mu_\sigma^+ + \sigma''z' - \sigma'q}{\sigma''z' - \sigma'q}\right)\right], \tag{22}$$

where we defined

$$F(q, z, \mu_\sigma) \equiv \frac{1}{8\mu_\sigma q}\left[(2\mu_\sigma + z)^2 - q^2\right]. \tag{23}$$

Eq. (22) is just the same result one gets for $t = 0$, but evaluated at $z' = z - \mathbf{q} \cdot \mathbf{t}$ and $\mu_\sigma^+ = \mu_\sigma/(1+t)$. For the second term we find

$$I_2(q, z, \mu_\sigma; t) - I_2(q, z, -\mu_\sigma; t) = 2q\mu_\sigma^+(q^2 - z'^2)\left[\frac{t(q - z')}{q + z't} + \sum_{\sigma', \sigma''} F(\sigma'q, \sigma''z', \mu_\sigma^+) \log\left(\frac{2\mu_\sigma^+ + \sigma''z' - \sigma'q}{\sigma''z' - \sigma'q}\right)\right.$$

$$- \frac{q}{8\mu_\sigma^+}\left\{\left(\frac{-2\mu_\sigma + qt + z'}{q + z't}\right)^2 - 1\right\}\left\{\log\left(\frac{2\mu_\sigma^- - z' - q}{-z' - q}\right) - \log\left(\frac{2\mu_\sigma^- - z' + q}{-z' + q}\right)\right\}$$

$$\left. + \frac{q}{8\mu_\sigma^+}\left\{\left(\frac{2\mu_\sigma + qt + z'}{q + z't}\right)^2 - 1\right\}\left\{\log\left(\frac{2\mu_\sigma^- + z' - q}{z' - q}\right) - \log\left(\frac{2\mu_\sigma^+ + z' + q}{z' + q}\right)\right\}\right]. \tag{24}$$

Finally, combining the results from Eq. (22) and Eq. (24) using Eq. (18) and reinstating $v_F$, we find

$$\boxed{\sigma_\perp^{\rm CME}(q, \omega) = \frac{iq_l \Pi_\perp^l(q, \omega^+; t)}{q^2} = \frac{e^2}{4\pi^2} \sum_{\sigma=\pm} \sigma\mu_\sigma \mathcal{S}_\perp^{\rm CME}\left(\frac{\omega^+}{v_F q} - t, \frac{\mu_\sigma}{v_F q}; t\right),} \tag{25}$$

with

$$\mathcal{S}_\perp^{\text{CME}}(x,y;t) = \frac{1-x^2}{2(1+xt)} - \sum_{\sigma,\sigma'=\pm} \sigma K_{\sigma'}(x,y;t) H_{\sigma\sigma'}(x,y;t), \tag{26}$$

in terms of the dimensionless functions

$$K_\sigma(x,y;t) \equiv \left(\frac{1-x^2}{16y}\right)\left[\left(\frac{2\sigma y+t+x}{1+xt}\right)^2 - 1\right] \quad \text{and} \quad H_{\sigma\sigma'}(x,y;t) \equiv \log\left(1 + \frac{2y}{(1-\sigma\sigma't)(\sigma'x-\sigma)}\right). \tag{27}$$

## IV. LONG-WAVELENGTH LIMIT CHIRAL MAGNETIC CONDUCTIVITY FOR ARBITRARY ANGLE TILT

In order to obtain the long-wavelength limit of the conductivities in Eq. (16), we take the limit $q \to 0$ while keeping $x \equiv z/v_F q$ fixed. The homogeneous limit then corresponds to $x \to \infty$, whereas the static limit correponds to $x \to 0$. To find the two conductivities separately, we need to calculate the projection of the current-current response function onto $\mathbf{t}$ and $\mathbf{q}$ and subsequently form the linear combinations as written down in Eq. (16). To calculate the necessary integrals we choose $\mathbf{q}$ along the $z$-axes without loss of generality and we denote $\mathbf{q}\cdot\mathbf{t}=qt_\parallel$. Furthermore, we can put $\mathbf{t}$ in the $xz$-plane. Hence, when going to spherical coordinates $(\phi,\theta,k)$, we have $\mathbf{q}\cdot\mathbf{k}=kq\cos\theta$ and $\mathbf{k}\cdot\mathbf{t}=k(t_\perp\cos\phi\sin\theta + t_\parallel\cos\theta)$. Again we use the fact that the Heaviside functions in Eq. (12) are only nonzero when $\sigma'\mu_\sigma > 0$.

As an example of the calculations to be done, we calculate the projection of $\Pi_l^{(1)}(\mathbf{q},z;\mathbf{t})$ onto $\mathbf{q}$. Note that the "(1)" refers to the part of the current-current response function which is not explicitly linear in $\mathbf{t}$. This means that we only include the first two pairs of terms in the form factor. Expanding for small $q$, we find when $\sigma' = +1$ and $\sigma'' = +1$,

$$\begin{aligned}
iq_l\Pi^{l(1)}_{++}(\mathbf{q},z;\mathbf{t}) &= \frac{e^2 v_F^2}{2}\sum_\sigma \sigma \int \frac{d^3\mathbf{k}}{(2\pi)^3}\left[\frac{-q^2\sin^2\theta(\cos\theta+t_\parallel)}{k(x'-\cos\theta)}\right]\delta(\mu_\sigma - v_F k - v_F \mathbf{k}\cdot\mathbf{t}) + \mathcal{O}(q^3)\\
&= -\frac{e^2 v_F^2 q^2}{2}\sum_\sigma \frac{\sigma}{(2\pi)^3}\int_0^{2\pi}d\phi\int_0^\pi d\theta\sin\theta\int_0^\infty dk\, k\left[\frac{\sin^2\theta(\cos\theta+t_\parallel)}{x'-\cos\theta}\right]\delta(\mu_\sigma - v_F k - v_F \mathbf{k}\cdot\mathbf{t}) + \mathcal{O}(q^3)\\
&= -\frac{e^2\mu_5}{(2\pi)^3}\int_0^{2\pi}d\phi\int_0^\pi d\theta\left[\frac{\sin^3\theta(\cos\theta+t_\parallel)}{x'-\cos\theta}\right]\left[\frac{1}{(1+t_\perp\cos\phi\sin\theta+t_\parallel\cos\theta)^2}\right] + \mathcal{O}(q^3)\\
&= -\frac{q^2 e^2\mu_5}{(2\pi)^2}\int_0^\pi d\theta\left[\frac{\sin^3\theta(\cos\theta+t_\parallel)(1+t_\parallel\cos\theta)}{[x'-\cos\theta]\left[(\cos\theta+t_\parallel)^2+(1-t^2)\sin^2\theta\right]^{3/2}}\right] + \mathcal{O}(q^3)\\
&= -\frac{q^2 e^2\mu_5}{(2\pi)^2}\int_{-1}^1 dy\left[\frac{(1-y^2)(y+t_\parallel)(1+t_\parallel y)}{[x'-y]\left[(y+t_\parallel)^2+(1-t^2)(1-y^2)\right]^{3/2}}\right] + \mathcal{O}(q^3),
\end{aligned} \tag{28}$$

where we abbreviated $x' \equiv z/v_F q - (\mathbf{q}\cdot\mathbf{t})/q \equiv x - (\mathbf{q}\cdot\mathbf{t})/q$ and used $2\mu_5 = \mu_+ - \mu_-$. The last integral can be done analytically, but we refrain from showing this partial result explicitly. Similarly, the contributions for $\sigma'=1, \sigma''=-1$ and $\sigma'=-1, \sigma''=1$ yield when we expand for small $q$,

$$\begin{aligned}
iq_l\left[\Pi^{l(1)}_{+-}+\Pi^{l(1)}_{-+}\right](\mathbf{q},z;\mathbf{t}) &= \frac{e^2 v_F^2}{2}\sum_\sigma \sigma \int \frac{d^3\mathbf{k}}{(2\pi)^3}\left[\frac{-q^2(x'\cos\theta+\cos 2\theta)}{v_F k^2}\right]\vartheta(\mu_\sigma - v_F k - v_F\mathbf{k}\cdot\mathbf{t}) + \mathcal{O}(q^3)\\
&= -\frac{q^2 e^2\mu_5}{(2\pi)^2}\int_0^{2\pi}d\phi\int_0^\pi d\theta\left[\frac{\sin\theta(x'\cos\theta+\cos 2\theta)}{1+t_\perp\cos\phi\sin\theta+t_\parallel\cos\theta}\right] + \mathcal{O}(q^3)\\
&= -\frac{q^2 e^2\mu_5}{(2\pi)^2}\int_0^\pi d\theta\left[\frac{\sin\theta(x'\cos\theta+\cos 2\theta)}{\sqrt{(\cos\theta+t_\parallel)^2+(1-t^2)\sin^2\theta}}\right] + \mathcal{O}(q^3)\\
&= -\frac{q^2 e^2\mu_5}{(2\pi)^2}\int_{-1}^1 dy\left[\frac{x'y+2y^2-1}{\sqrt{(y+t_\parallel)^2+(1-t^2)(1-y^2)}}\right] + \mathcal{O}(q^3).
\end{aligned} \tag{29}$$

Performing all remaining integrals explicitly and calculating the other contributions, we find for the two projections

$$iq_l\Pi^l(\mathbf{q},z;\mathbf{t}) = q^2(1+t_\parallel x' - t_\perp^2)\left[\frac{S^{\mathrm{CME}}(x-t_\parallel;\mathbf{t})}{1-t^2}\right]\frac{e^2\mu_5}{2\pi^2},$$

$$it_l\Pi^l(\mathbf{q},z;\mathbf{t}) = q(t_\parallel + t_\parallel^2 x' + t_\perp^2 x')\left[\frac{S^{\mathrm{CME}}(x-t_\parallel;\mathbf{t})}{1-t^2}\right]\frac{e^2\mu_5}{2\pi^2}, \tag{30}$$

where we defined

$$S^{\mathrm{CME}}(x';\mathbf{t}) \equiv \frac{(1-t^2)(1-x'^2)}{N^2(x';\mathbf{t})}\left[1 + \frac{(x'+t_\parallel)H(x';\mathbf{t})}{2N(x';\mathbf{t})}\right] = \begin{cases} 1 & \text{as} \quad x \to 0 \\ (1-t^2)l(t) & \text{as} \quad x \to \infty \end{cases}, \tag{31}$$

in terms of the functions

$$H(x';\mathbf{t}) \equiv \log\left(\frac{(x'-1)\left[1-t_\parallel + t_\parallel^2 + t_\parallel x' - t^2(1+x') + (1-t_\parallel)N(x';\mathbf{t})\right]}{(x'+1)\left[1+t_\parallel + t_\parallel^2 + t_\parallel x' - t^2(1-x') + (1+t_\parallel)N(x';\mathbf{t})\right]}\right), \tag{32}$$

and

$$N(x';\mathbf{t}) \equiv \sqrt{(1-t^2)(1-x^2) + (x+t_\parallel)^2}. \tag{33}$$

Now we are in the position to form the linear combinations from Eq. (16). We find

$$\sigma^{\mathrm{CME}}(\mathbf{q},\omega) = \left[\frac{q^2 t^2(1+t_\parallel x' - t_\perp^2) - q(\mathbf{q}\cdot\mathbf{t})(t_\parallel + t_\parallel^2 x' + t_\perp^2 x')}{q^2 t^2 - (\mathbf{q}\cdot\mathbf{t})^2}\right]\frac{e^2\mu_5}{2\pi^2}\frac{S^{\mathrm{CME}}(x';\mathbf{t})}{1-t^2}$$

$$= \boxed{\frac{e^2\mu_5}{2\pi^2}S^{\mathrm{CME}}\left(\frac{\omega^+}{v_F q} - t_\parallel;\mathbf{t}\right)}, \tag{34}$$

and

$$\sigma^{\mathrm{AHE}}(\mathbf{q},\omega) = \frac{1}{\omega}\left[\frac{q^2(\mathbf{q}\cdot\mathbf{t})\left(1+t_\parallel x' + \frac{t_\perp^2}{t_\parallel}x'\right) - (\mathbf{q}\cdot\mathbf{t})q^2(1+t_\parallel x' - t_\perp^2)}{q^2 t^2 - (\mathbf{q}\cdot\mathbf{t})^2}\right]\frac{e^2\mu_5}{2\pi^2}\frac{S^{\mathrm{CME}}(x';\mathbf{t})}{1-t^2}$$

$$= \boxed{\frac{e^2\mu_5}{2\pi^2 v_F(1-t^2)}S^{\mathrm{CME}}\left(\frac{\omega^+}{v_F q} - t_\parallel;\mathbf{t}\right)}. \tag{35}$$

Hence, as claimed in the main text we have $\sigma^{\mathrm{AHE}}(\mathbf{q},\omega) = \sigma^{\mathrm{CME}}(\mathbf{q},\omega)/v_F(1-t^2)$ in the long-wavelength limit.

## V. DYNAMICAL INTRINSIC ANOMALOUS HALL CONDUCTIVITY

When we project the response function in Eq. (12) onto $\mathbf{t}$ and consider $\mathbf{q} = \mathbf{0}$, the chiral magnetic conductivity drops out and we obtain the dynamical, planar, intrinsic anomalous Hall effect with current $\mathbf{J}(\omega) = \sigma^{\mathrm{AHE}}(\omega)\mathbf{t}\times\mathbf{E}(\omega)$. The corresponding dynamical anomalous Hall conductivity $\sigma^{\mathrm{AHE}}(\omega)$ is given by

$$\sigma^{\mathrm{AHE}}(\omega) = \frac{it^l\Pi_l(\mathbf{0},\omega^+;\mathbf{t})}{\omega t^2}. \tag{36}$$

To calculate the dynamical conductivity we project Eq. (12) onto $t_l$ and evaluate it at $\mathbf{q} = \mathbf{0}$, leading to

$$it^l\Pi_l(\mathbf{0},z;\mathbf{t}) = \frac{e^2 v_F^2}{2}\sum_{\sigma,\sigma',\sigma''}\sigma\int\frac{d^3\mathbf{k}}{(2\pi)^3}\frac{(\mathbf{t}\cdot\mathbf{k})(\sigma'-\sigma'')}{k}\left[\frac{\sigma'\vartheta(\sigma'\mu_\sigma(\mathbf{k})-E_\mathbf{k}) - \sigma''\vartheta(\sigma''\mu_\sigma(\mathbf{k})-E_\mathbf{k})}{z+\sigma' E_\mathbf{k} - \sigma'' E_\mathbf{k}}\right]$$

$$= -2ze^2 v_F^2 \sum_\sigma \sigma \int\frac{d^3\mathbf{k}}{(2\pi)^3}\left[\frac{\mathbf{t}\cdot\mathbf{k}}{k}\right]\left[\frac{\vartheta(\mu_\sigma - v_F\mathbf{k}\cdot\mathbf{t} - v_F|\mathbf{k}|)}{4v_F^2|\mathbf{k}|^2 - z^2} + \frac{\vartheta(-\mu_\sigma + v_F\mathbf{k}\cdot\mathbf{t} - v_F|\mathbf{k}|)}{4v_F^2|\mathbf{k}|^2 - z^2}\right], \tag{37}$$

where the first term is only nonzero for $\mu_\sigma > 0$, whereas the second term is only nonzero for $\mu_\sigma < 0$. The second term can be brought into the form of the first term by performing a change of integration variables $\mathbf{k} \to -\mathbf{k}$, at the cost of an overal minus sign. Below we find that the result of the integral over $\mathbf{k}$ of the first term is odd in $\mu_\sigma$. Thus, it suffices to calculate the first integral, yielding an answer valid for any sign of $\mu_\sigma$. We proceed by putting $\mathbf{t}$ in the $z$-direction without loss of generality and subsequently we use the spherical coordinates $(\phi, \theta, k)$. The azimuthal integration is trivial and yields $2\pi$. Thus, we find

$$\begin{aligned}\sigma^{\text{AHE}}(\omega) &= \frac{it^l \Pi_l(\mathbf{0}, \omega^+; \mathbf{t})}{\omega t^2} = -\frac{2e^2 v_F^2}{(2\pi)^2 t} \sum_\sigma \sigma \int_0^\pi d\theta \int_0^\infty dk \left[\frac{k^2 \sin\theta \cos\theta}{4v_F^2 k^2 - (\omega^+)^2}\right] \vartheta(\mu_\sigma - v_F k - v_F k t \cos\theta) \\
&= -\frac{2e^2}{(2\pi)^2 v_F t} \sum_\sigma \sigma \int_{-1}^1 dx\, x \int_0^{\frac{\mu_\sigma}{1+xt}} dk \left[\frac{k^2}{4k^2 - (\omega^+)^2}\right] \\
&= -\frac{e^2}{16(2\pi)^2 v_F t} \sum_\sigma \sigma \left[\frac{8\mu_\sigma}{t} + \left(\frac{(2\mu_\sigma + \omega^+)^2}{zt^2} - \omega^+\right)\left\{\log\left(1 + \frac{2\mu_\sigma}{(1+t)\omega^+}\right) - \log\left(1 + \frac{2\mu_\sigma}{(1-t)\omega^+}\right)\right\}\right. \\
&\qquad \left. + \left(\frac{(2\mu_\sigma - \omega^+)^2}{\omega^+ t^2} - \omega^+\right)\left\{\log\left(1 - \frac{2\mu_\sigma}{(1-t)\omega^+}\right) - \log\left(1 - \frac{2\mu_\sigma}{(1+t)\omega^+}\right)\right\}\right] \\
&= -\frac{e^2}{16(2\pi)^2 v_F t} \sum_\sigma \sigma \left[\frac{8\mu_\sigma}{t} + \frac{16\mu_\sigma}{t} L_+\left(\frac{\omega^+}{\mu_\sigma}; t\right)\left\{H_{-+}\left(0, \frac{\mu_\sigma}{\omega^+}; t\right) - H_{--}\left(0, \frac{\mu_\sigma}{\omega^+}; t\right)\right\}\right. \\
&\qquad \left. + \frac{16\mu_\sigma}{t} L_-\left(\frac{\omega^+}{\mu_\sigma}; t\right)\left\{H_{++}\left(0, \frac{\mu_\sigma}{\omega^+}; t\right) - H_{+-}\left(0, \frac{\mu_\sigma}{\omega^+}; t\right)\right\}\right] \\
&= \boxed{\frac{e^2}{(2\pi)^2 v_F} \sum_\sigma \sigma\mu_\sigma \left[-\frac{1}{2t^2} + \sum_{\sigma',\sigma''} \sigma'' L_{\sigma'}\left(\frac{\omega^+}{\mu_\sigma}; t\right) H_{\sigma'\sigma''}\left(0, \frac{\mu_\sigma}{\omega^+}; t\right)\right]},
\end{aligned} \qquad (38)$$

where we used the definition of $H_{\sigma\sigma'}(x, y; t)$ from Eq. (27) and defined the rational function

$$L_\sigma(y; t) \equiv \frac{y^2 t^2 - (2 - \sigma y)^2}{16 y t^3}. \qquad (39)$$

### Appendix A: Conventions

We use the Minkowski metric $\eta^{\mu\nu}$ with signature $(-+++)$. Fourier transforms are defined as

$$f(x) = \int \frac{d^4 k}{(2\pi)^4} f(k) e^{ik_\mu x^\mu} = \int \frac{d^4 k}{(2\pi)^4} f(k) e^{-i\omega t + i\mathbf{k}\cdot\mathbf{x}}. \qquad (A1)$$

The Clifford algebra reads $\{\gamma^\mu, \gamma^\nu\} = 2\eta^{\mu\nu}$. In terms of the 2x2 identity matrix $\mathbb{1}_2$ and the Pauli matrices $\sigma^i$, we represent our gamma matrices by

$$\gamma^0 = \begin{pmatrix} 0 & \mathbb{1}_2 \\ -\mathbb{1}_2 & 0 \end{pmatrix}, \quad \gamma^i = \begin{pmatrix} 0 & \sigma^i \\ \sigma^i & 0 \end{pmatrix} \quad \text{and} \quad \gamma^5 = \begin{pmatrix} -\mathbb{1}_2 & 0 \\ 0 & \mathbb{1}_2 \end{pmatrix}, \qquad (A2)$$

such that $(\gamma^0)^2 = -1$, $(\gamma^i)^2 = 1$ and $\{\gamma^5, \gamma^0\} = \{\gamma^5, \gamma^i\} = 0$.



\begin{thebibliography}{99}
\bibitem{Explain0}
	We do not consider a time-dependent chiral imbalance.
\bibitem{Explain1}
	One could also think of a term $\gamma(\bq, \w; \bt) \bq \times \bt$. However, from \eqqref{eq:pol} we see $\Pi_l(\bq,\w;\bt) = - \Pi_l(-\bq,\w;-\bt)$, implying $\gamma(\bq,\w;\bt) = - \gamma(-\bq,\w;-\bt)$. As $\gamma(\bq,\w;\bt)$ is a scalar, it must therefore be zero.
\bibitem{Explain2}
	Note that \eqqref{eq:cond1homg} can be rederived from \eqqref{eq:cond1} by expanding
    $( 1 + \w t/v_Fq(1-t^2))S^{\text{CME}}(\w/v_F q - t; \bt)$ up to first order in $\w/v_F q$. It is thus important to take into accound the terms of $O(v_Fq /\w)$ in the homogeneous limit of $S^{\text{CME}}(\w/v_F q - t; \bt)$.
\end{thebibliography}
\end{document}